\documentclass[twocolumn]{aastex62}
\usepackage{amsmath}
\usepackage{epsf}
\usepackage{color}
\usepackage{color}
\usepackage{enumitem}
\usepackage{mathtools}

\usepackage[mathscr]{euscript}
 \let\mathscr\relax
\usepackage[scr]{rsfso}

\newcommand\oiii{[{\sc O~iii}]}
\newcommand\oii{[{\sc O~ii}]}
\newcommand\cii{[{\sc C~ii}]}

\newcommand\nii{[{\sc N~ii}]}
\newcommand\oi{[{\sc O~i}]}
\newcommand\mgii{Mg~{\sc ii}}
\newcommand\civ{{\sc C~iv}}
\newcommand\sii{[{\sc S~ii}]}

\graphicspath{{./}{}}

\accepted{March 17, 2023}
\submitjournal{ApJ}

\shorttitle{FIR Lines of Mrk~54}
\shortauthors{Ura et al.}

\begin{document}

\title{Detections of \cii\ 158 \micron\ and \oiii\ 88 \micron\ in a Local Lyman Continuum Emitter, Mrk~54, and its Implications to High-redshift ALMA Studies\footnote{Based on observations done with FIFI-LS on SOFIA}}

\correspondingauthor{Takuya Hashimoto}
\email{hashimoto.takuya.ga@u.tsukuba.ac.jp}

\author{Ryota Ura}
\affiliation{Division of Physics, Faculty of Pure and Applied Sciences, University of Tsukuba,Tsukuba, Ibaraki 305-8571, Japan}
\author[0000-0002-0898-4038]{Takuya Hashimoto}
\affiliation{Division of Physics, Faculty of Pure and Applied Sciences, University of Tsukuba,Tsukuba, Ibaraki 305-8571, Japan}
\affiliation{Tomonaga Center for the History of the Universe (TCHoU), Faculty of Pure and Applied Sciences, University of Tsukuba, Tsukuba, Ibaraki 305-8571, Japan}
\author[0000-0002-7779-8677]{Akio K. Inoue}
\affiliation{Department of Physics, School of Advanced Science and Engineering, Waseda University, 3-4-1, Okubo, Shinjuku, Tokyo 169-8555}
\affiliation{Waseda Research Institute for Science and Engineering, Faculty of Science and Engineering, Waseda University, 3-4-1, Okubo, Shinjuku, Tokyo 169-8555, Japan}
\author[0000-0002-3698-7076]{Dario Fadda}
\affiliation{SOFIA Science Center, Universities Space Research Association, MS 232-11, Moffett Field, CA 94035, USA}
\author[0000-0001-8587-218X]{Matthew Hayes}
\affiliation{The Oskar Klein Centre, Department of Astronomy, Stockholm University, AlbaNova, SE-10691 Stockholm, Sweden}
\author[0000-0003-1111-3951]{Johannes Puschnig}
\affiliation{The Oskar Klein Centre, Department of Astronomy, Stockholm University, AlbaNova, SE-10691 Stockholm, Sweden}
\affiliation{Argelander-Institut f\"{u}r Astronomie, Auf dem H\"{u}gel 71, D-53121, Bonn, Germany}
\author[0000-0003-1096-2636]{Erik Zackrisson}
\affiliation{Observational Astrophysics, Department of Physics and Astronomy, Uppsala University, Box 516, SE-751 20 Uppsala, Sweden}
\author[0000-0003-4807-8117]{Yoichi Tamura}
\affiliation{Department of Physics, Graduate School of Science, Nagoya University, Nagoya 464-8602, Japan}
\author[0000-0003-3278-2484]{Hiroshi Matsuo}
\affiliation{National Astronomical Observatory of Japan,
2-21-1 Osawa, Mitaka, Tokyo 181-8588, Japan}
\affiliation{Graduate University for Advanced Studies (SOKENDAI), 2-21-1 Osawa, Mitaka, Tokyo 181-8588, Japan}
\author[0000-0003-4985-0201]{Ken Mawatari}
\affiliation{National Astronomical Observatory of Japan, 2-21-1 Osawa, Mitaka, Tokyo 181-8588, Japan}

\author[0000-0001-7440-8832]{Yoshinobu Fudamoto}
\affiliation{Waseda Research Institute for Science and Engineering, Faculty of Science and Engineering, Waseda University, 3-4-1, Okubo, Shinjuku, Tokyo 169-8555, Japan}
\affiliation{National Astronomical Observatory of Japan, 2-21-1 Osawa, Mitaka, Tokyo 181-8588, Japan}
\author[0000-0001-8083-5814]{Masato Hagimoto}
\affiliation{Department of Physics, Graduate School of Science, Nagoya University, Nagoya 464-8602, Japan}
\author[0000-0002-1234-8229]{Nario Kuno}
\affiliation{Division of Physics, Faculty of Pure and Applied Sciences, University of Tsukuba,Tsukuba, Ibaraki 305-8571, Japan}
\affiliation{Tomonaga Center for the History of the Universe (TCHoU), Faculty of Pure and Applied Sciences, University of Tsukuba, Tsukuba, Ibaraki 305-8571, Japan}
\author[0000-0001-6958-7856]{Yuma Sugahara}
\affiliation{Waseda Research Institute for Science and Engineering, Faculty of Science and Engineering, Waseda University, 3-4-1, Okubo, Shinjuku, Tokyo 169-8555, Japan}
\affiliation{National Astronomical Observatory of Japan, 2-21-1 Osawa, Mitaka, Tokyo 181-8588, Japan}
\author[0000-0002-7738-5290]{Satoshi Yamanaka}
\affiliation{General Education Department, National Institute of Technology, Toba College, 1-1, Ikegami-cho, Toba, Mie 517-8501, Japan}
\author[0000-0002-5268-2221]{Tom J. L. C. Bakx}
\affiliation{Department of Physics, Graduate School of Science, Nagoya University, Nagoya 464-8602, Japan}
\affiliation{National Astronomical Observatory of Japan, 2-21-1, Osawa, Mitaka, Tokyo, Japan}
\author[0000-0002-0984-7713]{Yurina Nakazato}
\affiliation{Department of Physics, School of Science, The University of Tokyo, 7-3-1 Hongo, Bunkyo, Tokyo 113-0033, Japan}
\author{Mitsutaka Usui}
\affiliation{Division of Physics, Faculty of Pure and Applied Sciences, University of Tsukuba,Tsukuba, Ibaraki 305-8571, Japan}
\author[0000-0002-1319-3433]{Hidenobu Yajima}
\affiliation{Division of Physics, Faculty of Pure and Applied Sciences, University of Tsukuba,Tsukuba, Ibaraki 305-8571, Japan}
\affiliation{Center for Computational Sciences, University of Tsukuba, Ten-nodai, 1-1-1 Tsukuba, Ibaraki 305-8577, Japan}
\author[0000-0001-7925-238X]{Naoki Yoshida}
\affiliation{Department of Physics, School of Science, The University of Tokyo, 7-3-1 Hongo, Bunkyo, Tokyo 113-0033, Japan}
\affiliation{Kavli Institute for the Physics and Mathematics of the Universe (WPI), UT Institute for Advanced Study, The University of Tokyo, Kashiwa, Chiba 277-8583, Japan}

\begin{abstract}
We present integral field, far-infrared (FIR) spectroscopy of Mrk~54, a local Lyman Continuum Emitter (LCE), obtained with FIFI-LS on the Stratospheric Observatory for Infrared Astronomy. This is only the second time, after Haro~11, that \cii~158~\micron\ and \oiii~88~\micron\ spectroscopy of the known LCEs have been obtained.
We find that Mrk~54 has a strong \cii\ emission that accounts for $\sim1$\% of the total FIR luminosity, whereas it has only moderate \oiii\ emission, resulting in the low \oiii/\cii\ luminosity ratio of $0.22\pm0.06$.
In order to investigate whether \oiii/\cii\ is a useful tracer of $f_{\rm esc}$ (LyC escape fraction), we examine the correlations of \oiii/\cii\ and (i) the optical line ratio of $\rm O_{32}$~$\equiv$~\oiii~5007~\AA/\oii~3727~\AA, (ii) specific star formation rate, (iii) \oiii~88~\micron/\oi~63~\micron\ ratio, (iv) gas phase metallicity, and (v) dust temperature based on a combined sample of Mrk~54 and the literature data from the {\it Herschel} Dwarf Galaxy Survey and the LITTLE THINGS Survey. We find that galaxies with high \oiii/\cii\ luminosity ratios could be the result of high ionization (traced by $\rm O_{32}$), bursty star formation, high ionized-to-neutral gas volume filling factors (traced by \oiii~88~\micron/\oi~63~\micron), and low gas-phase metallicities, which is in agreement with theoretical predictions. 
We present an empirical relation between the \oiii/\cii\ ratio and $f_{\rm esc}$ based on the combination of the \oiii/\cii\ and $\rm O_{32}$ correlation, and the  known relation between $\rm O_{32}$ and $f_{\rm esc}$. 
The relation implies that high-redshift galaxies with high \oiii/\cii\ ratios revealed by ALMA may have $f_{\rm esc}\gtrsim0.1$, significantly contributing to the cosmic reionization. 
\end{abstract}
\keywords{galaxies: ISM, galaxies: starburst, infrared: galaxies}

\section{Introduction} \label{sec:intro}

Studying high redshift ($z\gtrsim6$) galaxies at the epoch of reionzation (EoR) is essential to understand the cosmic reionization. An important parameter of galaxies necessary to understand the cosmic reionization is the Lyman continuum (LyC) escape fraction, $f_{\rm esc}$, which indicates the fraction of ionizing photons ($\lambda \le 912$ \AA) in a galaxy that escape from the galaxy into the intergalactic medium, IGM (e.g., \citealt{inoue2006,robertson2013, robertson2021}). 

Numerous efforts have been made to directly observe LyC. Space telescopes such as {\it FUSE} and {\it HST} identified LyC in more than {\bf ~$\sim100$} galaxies in the local and nearby Universe at $z\lesssim0.4$, and measured $f_{\rm esc}$ ranging from a few to more than 70\% (e.g., \citealt{deharveng2001, bergvall2006, leitet2011,leitet2013, borthakur2014, leitherer2016, izotov2016a, izotov2016b, izotov2018, puschnig2017, wang2019, flury2022a}). 
At higher redshift, spectroscopic detections of LyC in individual galaxies are limited so far. An AstroSat spectroscopically identified LyC in a $z=1.42$ galaxy, whose $f_{\rm esc}$ is constrained to be higher than 20\% (\citealt{kanak2020}). Large ground-based telescopes such as Keck, VLT and GTC spectroscopically identified LyC in $\sim 20$ individual galaxies at $z\sim3-4$ with $f_{\rm esc}$ ranging from a few to 90\% (e.g., \citealt{shapley2016, de_barros2016, vanzella2018, pahl2021, Marques-Chaves2021, Marques-Chaves2022}). 
Complementary to these individual spectroscopic detections, there are spectroscopic and imaging surveys at $z\sim3-4$ that statistically targeted the LyC radiation from galaxies. On the spectroscopy side, the Keck Lyman Continuum Spectroscopic Survey targeted LyC from 124 galaxies at $z\sim3.1$, and obtained a sample-average $f_{\rm esc}$ of $6\pm1$\%\ (\citealt{steidel2018, pahl2021}). On the imaging side, using narrow-band filters that are free from non-ionizing photons, one can observe LyC from galaxies with large ground-based telescopes such as Subaru, VLT, and Keck (e.g., \citealt{inoue2005, iwata2009, nestor2011, micheva2017, iwata2019}). Compared to spectroscopic surveys, narrow-band imaging observations can efficiently target LyC owing to large field-of-views. However, ground-based imaging data can be affected by chance overlap with foreground interlopers (e.g., \citealt{vanzella2010, nestor2013}). To mitigate such interlopers, \cite{iwata2019} combined the narrow-band imaging data of Subaru with the high angular resolution imaging data of {\it HST} data. The authors obtained $f_{\rm esc} < 8$\% ($3\sigma$) in a sample of 103 and 8 star-forming galaxies and AGNs with spectrsoscopic redshifts at $3.06 < z < 3.5$ and 157 photometrically-selected $z=3.1$ Ly$\alpha$-emitting galaxies. 
In addition to the narrow-band imaging surveys, \cite{fletcher2019} and \cite{begley2022} used broad-band filters to target LyC in samples of 61 and 148 galaxies at $z\sim3.1$ and $3.5$, respectively, and obtained the average $f_{\rm esc}$ of 20\% and $7\pm2$\% (see also, e.g., \citealt{jones2021, saxena2022.lyc, rivera-thorsen2022}).
At $z>4$, it is virtually impossible to directly observe LyC due to absorption by the intervening intergalactic medium, IGM (\citealt{inoue_iwata2008}). Therefore, it becomes important to establish a link between $f_{\rm esc}$ and other observables at $z=0-4$, which allows us to indirectly estimate how EoR galaxies have contributed to reionization. 

Possible signatures of LyC leakage include (i) a small separation of a double-peaked Ly$\alpha$ profile (e.g., \citealt{verhamme2015, dijkstra2016, kakiichi2021}), (ii) intense far ultraviolet emission lines such as Ly$\alpha$ (e.g., \citealt{steidel2018, naidu2022, flury2022b}), \civ\ 1550 \AA\ (e.g., \citealt{schaerer2022, saxena2022.civ}), and \mgii\ doublet lines at $\lambda \lambda 2796, 2803$ (e.g., \citealt{henry2018, chisholm2020}), (iii) a high ionization parameter traced by the optical line ratio \oiii~5007~\AA/\oii~3727~\AA\ ($\rm O_{32}$, e.g., \citealt{jaskot_oey2013, nakajima_ouchi2014}), (iv) a high starformation rate (SFR) surface density (\citealt{flury2022b}), (v) a blue ultraviolet (UV) continuum slope (e.g., \citealt{yamanaka2020, Chisholm2022}), (vi) and a weak equivalent width of low-ionization metal absorption lines (e.g., \citealt{Chisholm2018, Jones2013}) as well as a weak low-ionization optical metal emission lines such as \sii\ at $\lambda \lambda 6717, 6731$ (\citealt{wang2019}). On the whole, the LyC escape appears depend on H~{\sc i} column density, ionization parameter, and stellar feedback \citep{flury2022b}, as also predicted by theoretical studies (see, e.g., \citealt{yajima2011, yajima2014, zackrisson2013, Ma2015, Paardekooper2015, Arata2019, Barrow2020, Yeh2022, Rosdahl2022}).

In addition to these techniques, \citet{katz2020,katz2022} propose the far-infrared (FIR) line luminosity ratios of \oiii~88~\micron\ and \cii~158~\micron\ (hereafter, \oiii/\cii) as a new diagnostic signature of LyC escape. This is motivated by spectroscopic observations of $z=6-9$ galaxies with ALMA (\citealt{inoue2016, laporte2017, hashimoto2018, tamura2019}). \cite{inoue2016} observed a $z=7.2$ Ly$\alpha$-emitting galaxy in \oiii~88~\micron\ and \cii~158~\micron, and obtained a high line luminosity ratio \oiii/\cii~$\gtrsim12$ ($3\sigma$). With a simple one-dimensional photoionization model, the authors for the first time proposed that the high line ratio may indicate a high $f_{\rm esc}$ because a weak \cii\ emission could be linked to a low amount of H~{\sc i} gas in the interstellar medium (ISM). \cite{hashimoto2019a} compiled a sample of galaxies with \oiii/\cii\ at $z>6$, and found that UV-selected galaxies in the EoR typically show  \oiii/\cii~$\sim 3-10$ or higher (see also \citealt{laporte2019, bakx2020, harikane2020, carniani2020, witstok2022}). These values are higher than the typical values of local spirals ($\sim 0.5$; \citealt{brauher2008}) and low-metallicity dwarf galaxies ($\sim2$; \citealt{madden2013, cormier2015}) observed by {\it ISO} and {\it Herschel}, respectively, suggesting unusual conditions of the ISM in high-$z$ galaxies (e.g., \citealt{Sugahara2022}). 

Using the photoionization models of CLOUDY constructed by \citet{ferland2013}, \citet{harikane2020} showed that these high \oiii/\cii\ ratios can be explained by combinations of various parameters such as a high ionization parameter, low neutral gas covering fraction in the ISM, low gas density, and high O/C abundance ratio, where the scenarios of high ionization parameter and low neutral gas covering fraction would facilitate the LyC escape. In addition, \cite{vallini2021} have theoretically parameterized the \oiii/\cii\ ratio as a function of burstiness of star formation activity, gas density, and metallicity. The authors found that the parameter mostly affecting the line ratio is the burstiness. More recently, \cite{ramabason2022} examined ISM properties of a local sample of the \textit{Herschel} Dwarf Galaxy Survey (HDGS, \citealt{madden2013, cormier2015}) using a grid of models of  {\sc H~ii} regions and photo-dissociated regions. The authors have also found a positive correlation between \oiii/\cii and predicted $f_{\rm esc}$.
Although many parameters are degenerate, the fact that the EoR UV-selected galaxies show on average high \oiii/\cii\ luminosity ratios is interesting because it may suggest high $f_{\rm esc}$ values. It is crucial to observationally link the luminosity ratios to $f_{\rm esc}$, as done with other techniques, as well as to understand the characteristics of galaxies with high \oiii/\cii\ ratios.  If such correlations are established in the local Universe, we can apply them to the EoR galaxies observed with ALMA to indirectly infer their properties, including $f_{\rm esc}$. Unfortunately, only for one known local Lyman Continuum Emitter (LCE), Haro~11 (e.g., \citealt{hayes2007, leitet2011}), the \oiii/\cii\ ratio has been observed ($\sim2.0$ in \citealt{cormier2015}). The reason for the paucity of local LCEs with FIR line observations is that most LCEs were discovered after 2013, when {\it Herschel} ended its operations. 

The Stratospheric Observatory for Infrared Astronomy (SOFIA) offered a unique opportunity to perform FIR line spectroscopy of local galaxies. Here, we report observations of the FIR lines in Mrk~54 using FIFI-LS (Field-Imaging Far-Infrared Line-Spectrometer, \citealt{fischer2018}) onboard SOFIA; Mrk~54 is the second nearest galaxy after Haro~11 with a direct LyC detection at a luminosity distance of $\sim 191$ Mpc. The galaxy is a blue compact galaxy with $12+\log($O/H$) = 8.6$, SFR$ = 12~M_{\rm \odot}$~yr$^{-1}$, $f_{\rm esc}=2.5$\%\ (\citealt{leitherer2016}) and $\rm O_{32}$ = 0.4 (\citealt{Chisholm2018}). These two well-characterized LCEs with FIR line spectroscopy, Haro~11 and Mrk~54, are invaluable to calibrate an observational relationship between \oiii/\cii\ and $f_{\rm esc}$. Because \oiii/\cii\ depends on various parameters, we also make use of local analogues of high-$z$ galaxies with FIR line spectroscopy (e.g., low stellar mass or gas-phase metallicity), the HDGS and the LITTLE THINGS Survey (LT, \citealt{hunter2012, cigan2016, cigan2021}), to statistically examine the correlations between \oiii/\cii\ and other physical quantities such as ionization parameter, ionized-to-neutral gas volume filling factor, gas-phase metallicity, and burstiness. 
In light of recent combined observations of high-$z$ galaxies with ALMA and the {\it James Webb Space Telescope (JWST)} in the rest-frame optical and FIR (e.g., \citealt{bakx2023}), respectively, the presented study will be a reference study that is useful (1) to infer $f_{\rm esc}$ of galaxies in the EoR and (2) to plan and interpret the ALMA+JWST observations (e.g., \citealt{nakazato2023, kohandel2023}).

This paper is structured as follows. In \S~2, we present the SOFIA observations and data reduction of Mrk~54. In \S~3, we summarize the physical properties of the two literature samples of HDGS and LT. In \S~4, we present analyses of statistical correlations between \oiii/\cii\ and other physical quantities. In \S~5, we establish a relation between  \oiii/\cii\ and $f_{\rm esc}$, and apply it to high-$z$ ALMA galaxies. We summarize our conclusions in \S~6. Throughout this paper, we assume a $\Lambda$CDM cosmology with $\Omega_{\small m} = 0.272$, $\Omega_{\small b} = 0.045$, $\Omega_{\small \Lambda} = 0.728$ and $H_{\small 0} = 70.4$ km s$^{-1}$ Mpc$^{-1}$ (\citealt{komatsu2011}). We use $L_{\rm \odot}=3.839\times10^{33}$ erg s$^{-1}$ as solar luminosity and 12+log(O/H) = 8.7 as solar metallicity (\citealt{Asplund2009}).

\section{SOFIA Observations and Data Reduction} \label{sec:sofia_obs}

\begin{figure*}[]
\begin{center}
\hspace*{-1cm}
\includegraphics[width=8cm]{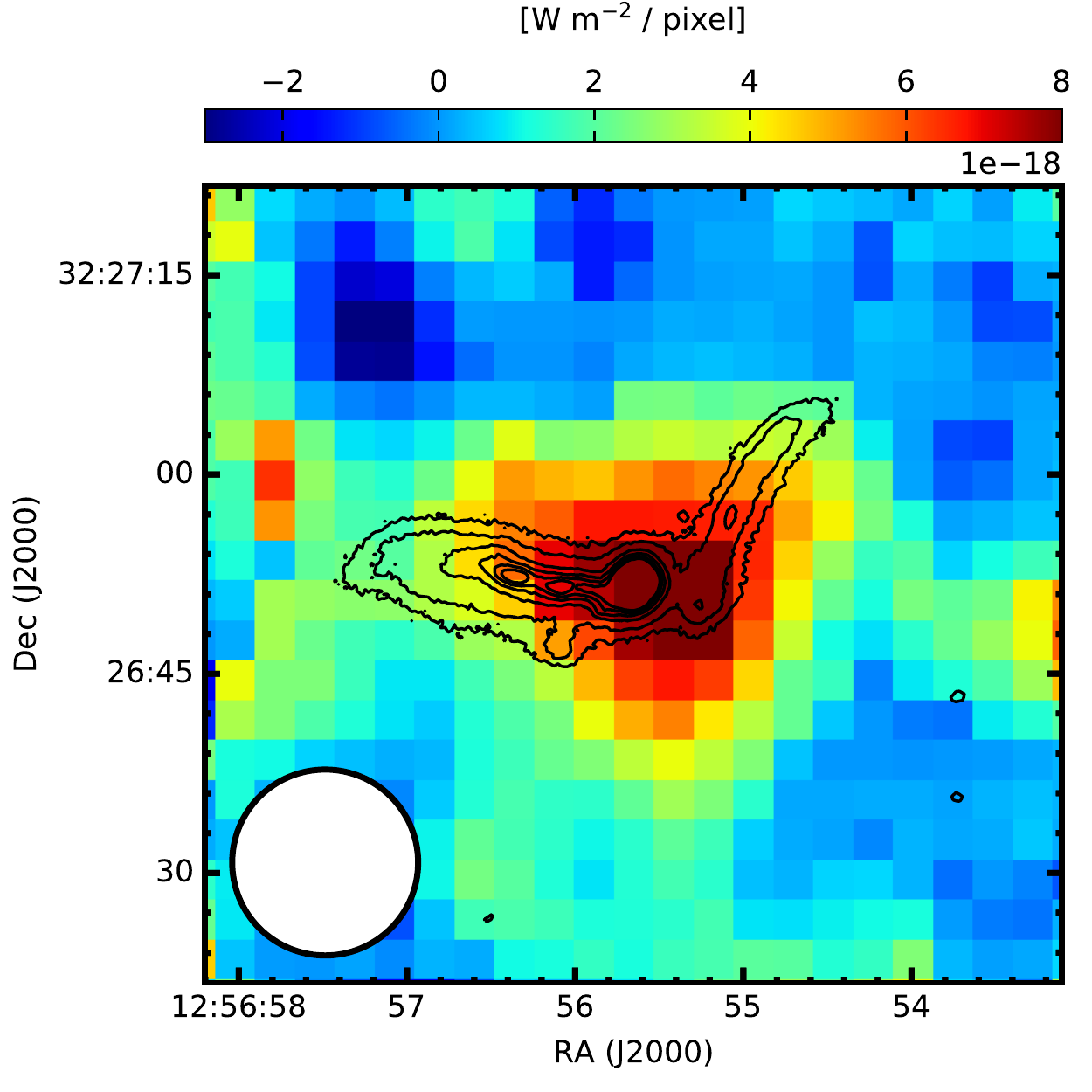}
\includegraphics[width=8cm]{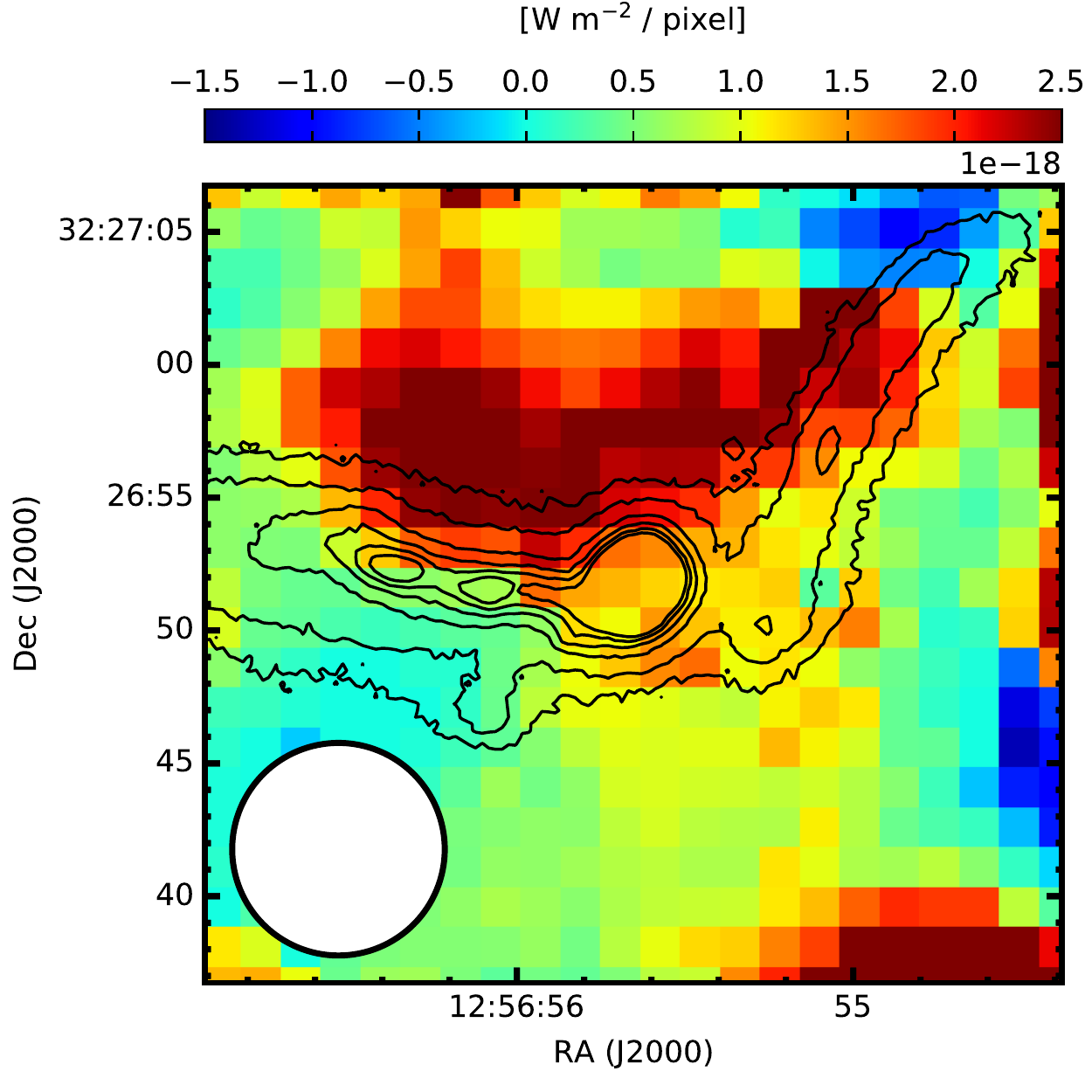}
\vspace*{-2cm}
\includegraphics[width=8cm]{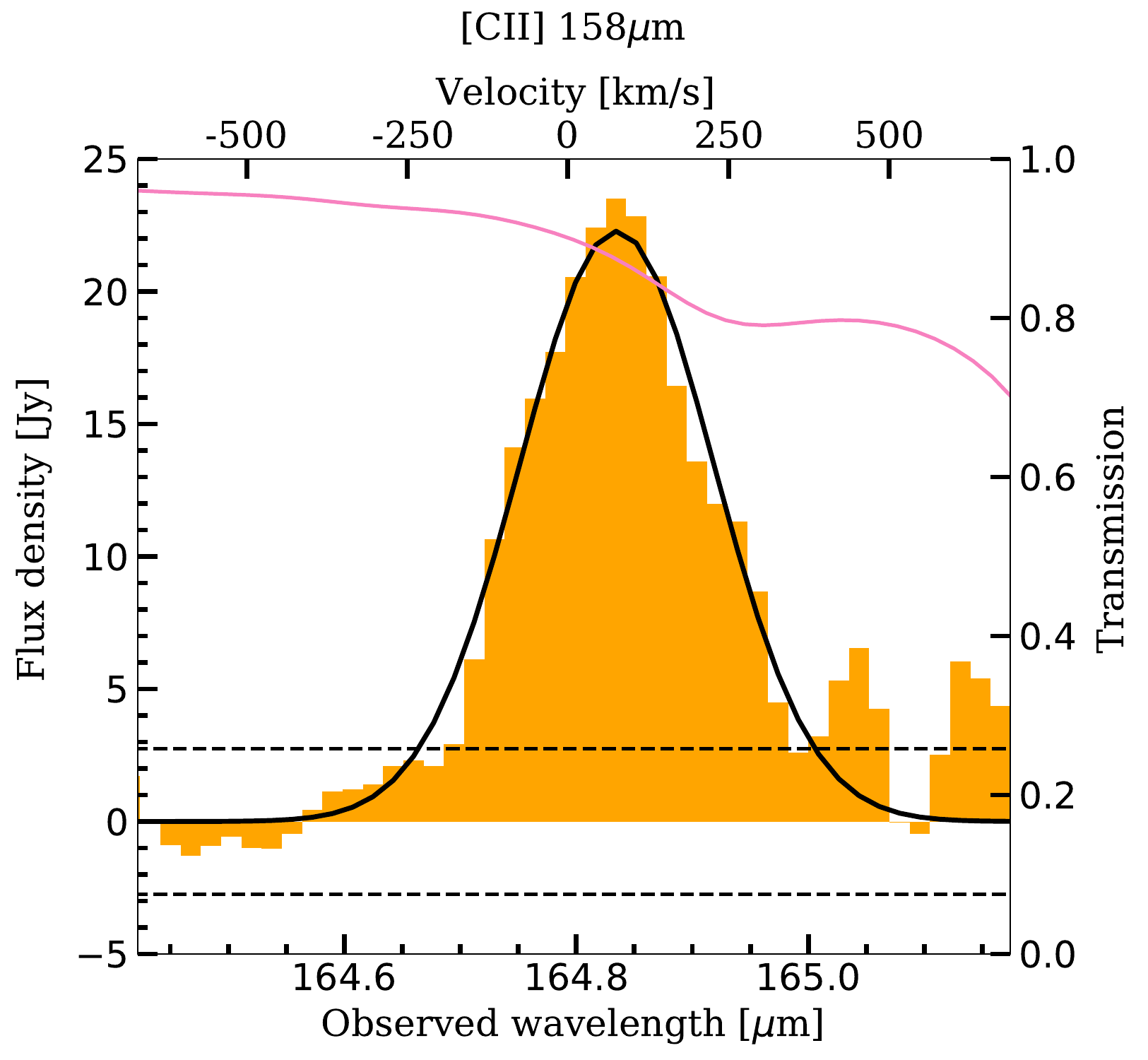}
\includegraphics[width=8cm]{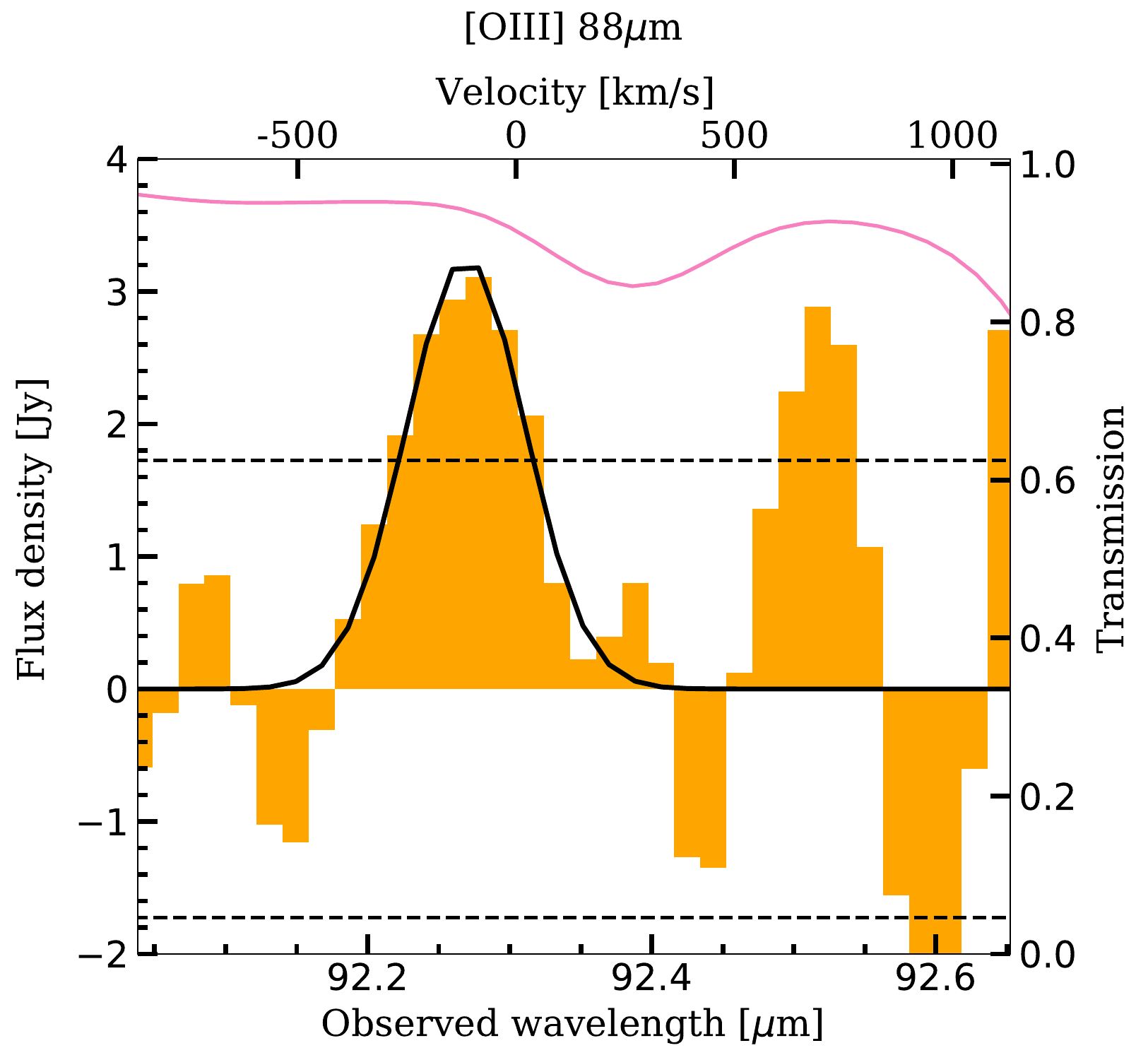}
\vspace*{2cm}
\end{center}
\caption{
Top left and right panels show the intensity maps of \cii\ and \oiii\ as the colored background, respectively, with flux scales denoted in the scale bar. In each panel, the black contours depict the morphology of Mrk~54 in the Pan-STARRS $i$-band image. Note that the left and right panels do not have the same physical scale because of the different field-of-views. The circles at the bottom left corners indicate the beam size of FIFI-LS.
Bottom left and right panels show the spectra of \cii\ and \oiii, respectively. In each panel, the magenta curve indicates the atmospheric transmission curve convolved with the spectral resolution of FIFI-LS. The black solid line shows the best-fit Gaussian profile to the line, whereas the horizontal black dashed line shows the typical $1\sigma$ uncertainty. In the top $x$-axis, the velocity zero-point is defined at $z=0.0447$ (\citealt{deharveng2001}).
\label{fig:mrk54}}
\end{figure*}

\begin{figure*}[]
\begin{center}
\hspace*{-1cm}
\includegraphics[width=19cm]{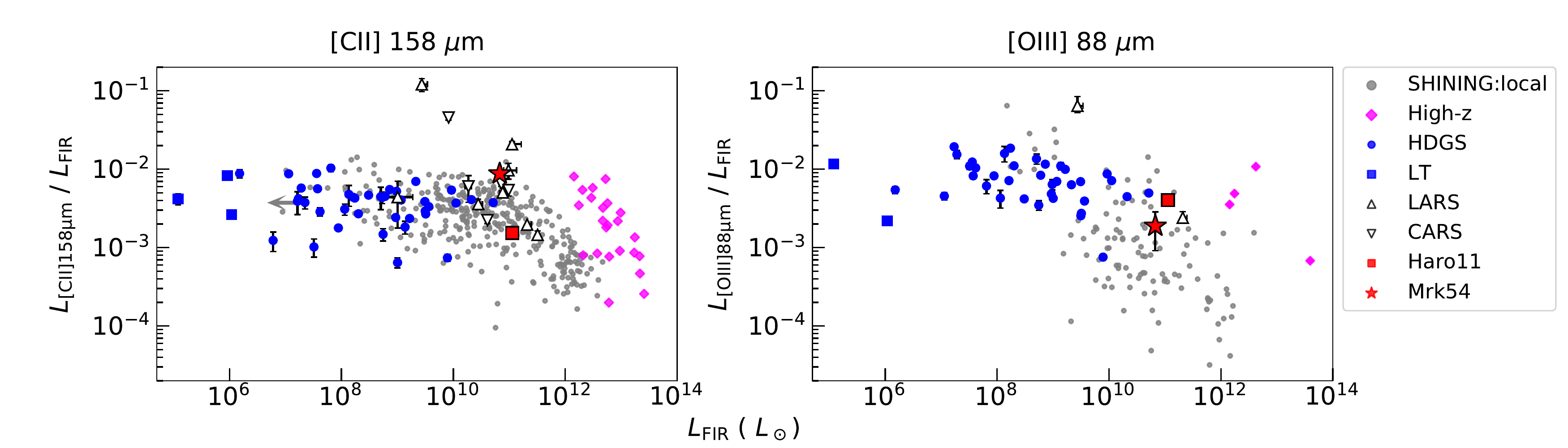}
\end{center}
\caption{
Left and right panels show \oiii- and \cii-to-FIR line luminosity ratios, $L_{\rm [CII]}/L_{\rm FIR}$ and $L_{\rm [OIII]}/L_{\rm FIR}$ plotted against $L_{\rm FIR}$, respectively. In each panel, Mrk~54 and Haro~11 are indicated by a red star and square with measurement uncertainties, respectively. Grey circles and magenta diamonds are the compilation of the FIR line observations of local and high-$z$ objects, respectively, from \cite{herrera-camus2018}, where the local sample includes {\sc H~ii} galaxies, LINER, and Seyfert. The blue circles and squares show the data points of HDGS (\citealt{cormier2012}) and LT (\citealt{cigan2016}). The upward and downward triangles indicate a sub-sample of local galaxies from the LARS (\citealt{puschnig2020}) and  CARS projects (\citealt{Smirnova-Pinchukova2019}) with the FIR line observations, some of which are characterized by enhanced FIR emission.} Mrk~54 has strong \cii\ emission, whereas it has moderate \oiii\ emission. 
\label{fig:Lline_Lratio}
\end{figure*}

\begin{figure}[]
\begin{center}
\vspace*{1cm}
\includegraphics[width=7cm]{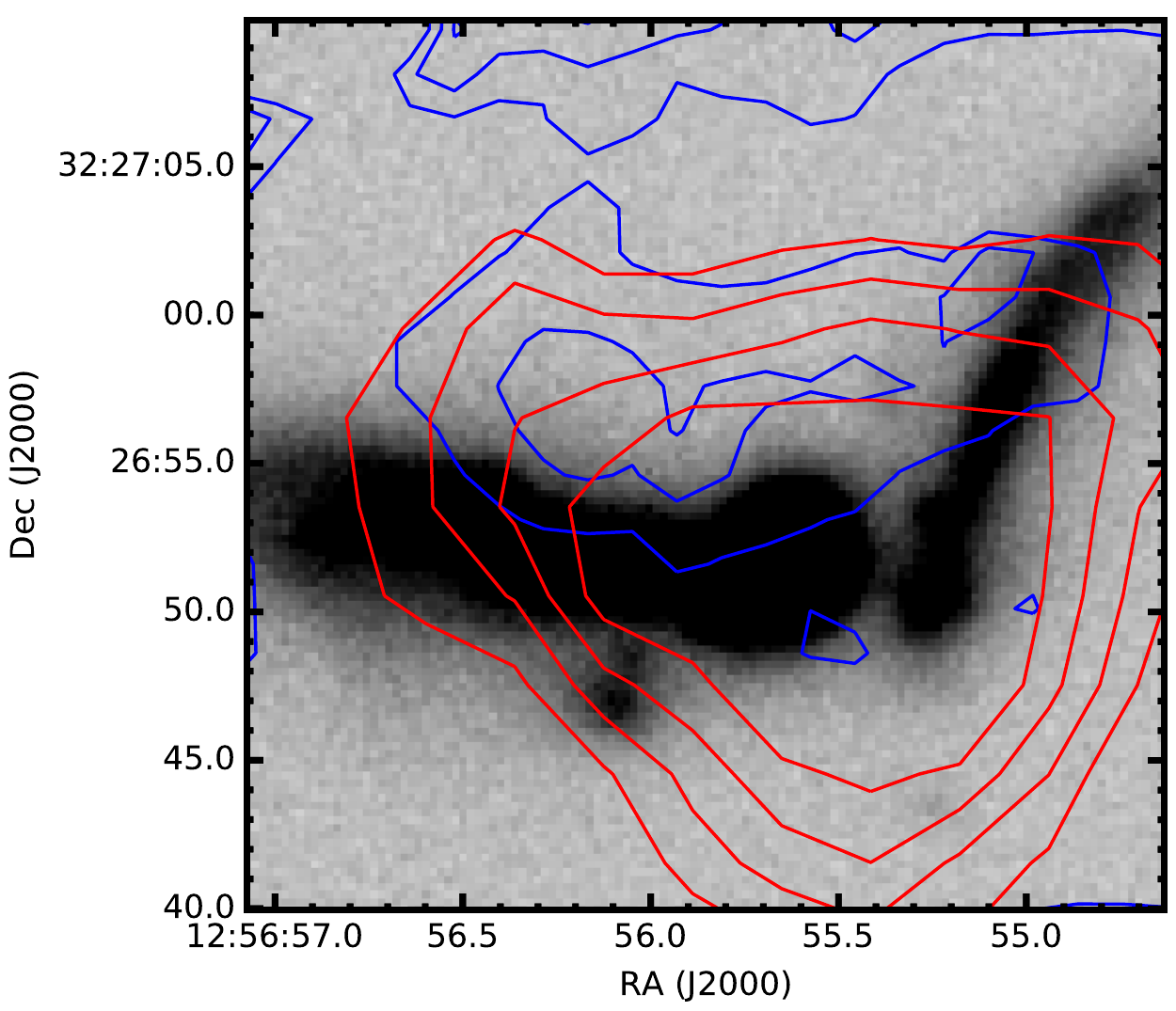}
\end{center}
\caption{
Comparisons of the spatial distributions of \cii, \oiii, and the Pan-STARRS $i$-band image. Red and blue contours indicate the spatial distributions of \cii\ and \oiii, respectively, with the Pan-STARRS $i$-band image as the background. The contours are drawn at an equal intervals with an arbitrary unit for the display purpose.
\label{fig:oiii_cii_panstarrs}}
\end{figure}

Mrk~54 was observed by the SOFIA Far Infrared Field-Imaging Line Spectrometer (FIFI-LS; \citealt{fischer2018}) during cycles 7 and 8 (PIDs: 07\_0108 and 08\_0063, PI: T. Hashimoto). FIFI-LS has two parallel spectral channels that cover $1' \times 1'$ in the wavelength ranges of $115-203$ \micron\ (``red channel'') and $0\farcm5 \times 0\farcm5$ in the ranges of $51-125$ \micron\ (``blue channel''), split into $5\times5$ spatial pixels. The red and blue channels were used to target the \cii~158~\micron\ and \oiii~88~\micron\ lines. The two observations were executed on Feb 2020 and May 2021 for a total on-source exposure time of 2426 seconds. The target galaxy Mrk~54 was entirely covered by a single field-of-view. The observations were performed in symmetric chopping mode with a $120\arcsec$ chop amplitude. A spatial dithering pattern was used to cover regions with bad pixels and to better recover the spatial resolution of the instrument. Eight grating positions were used to slightly extend the spectral coverage and to oversample the wavelength range in order
to compensate for bad pixels and to better define the line profiles\footnote{We only use wavelengths observed at 80\% or more of the time relative to the total observation time.}.

At the wavelengths of \cii~158~\micron\ and \oiii~88~\micron\ lines, the spatial resolution is approximately $15.6\arcsec$ and $8.8\arcsec$, the instantaneous spectral coverage is 1560 and 2340~km~s$^{-1}$, and the spectral resolution is 260 and $490~\mathrm{km~s^{-1}}$. Further details on FIFI-LS can be found in \cite{fischer2018} and \cite{colditz2020}. 

Data reduction was carried out using the FIFI-LS data reduction pipeline (\citealt{vacca2020}) combining the two sets of data taken in cycles 7 and 8. The flux calibration, which is performed using sky calibrators, has an absolute uncertainty less than $20$\%. Hereafter, we only consider the measurement uncertainties. The data were corrected for the atmospheric absorption with a transmission curve modeled with ATRAN \citep{Lord1992}, using precipitable water vapor values measured during the flight. The final cubes were sampled on a grid with spatial pixels of 3\arcsec\ and 1.5\arcsec\ (a fourth of the original pixels) and spectral pixels of 32.5~km~s$^{-1}$ and 61.25~km~s$^{-1}$ (oversample of a factor 8 in spectral resolution) for the red and blue array, respectively.

The data were analyzed with the SOFIA SPectral EXplorer ({\it sospex}, \citealt{fadda2018}). We first obtained a pure line cube by subtracting the dust continuum level from the spectral cube at each spatial pixel. We then defined an elliptical aperture by coadding the cube along the wavelength dimension and considering ellipses centered on the flux centroid of this image. Major and minor axes were optimized by selecting the smallest ellipse containing the asymptotic value of total flux.
Finally, we extracted the spectrum within this elliptical aperture and fit a Gaussian function to the line to estimate total flux, central wavelength, and full width at half maximum (FWHM).

The top left (right) panel of Figure \ref{fig:mrk54} shows the \cii\ (\oiii) intensity map with contours of the Pan-STARRS $i$-band image, whereas the bottom left (right) panel depicts the corresponding \cii\ (\oiii) spectrum extracted from the aperture. The \cii\ (\oiii) line is detected at a significance level of $24\sigma$ ($4\sigma$) integrated over the line profile, where $1\sigma$ corresponds to the typical uncertainty per spectral bin measured in the spectrum as indicated by the horizontal dashed lines\footnote{We measured the noise level per spectral bin, and adopted the standard deviation of the distribution as the $1\sigma$ value. We used the wavelength ranges of [164.42, 164.60] and [165.02, 165.17] \micron\ for \cii, whereas [92.04, 92.15] and [92.43, 92.65] \micron\ for \oiii. In this estimates, we neglected the wavelength dependence of the noise level for simplicity.}. 

Motivated by recent observations of high-$z$ galaxies that demonstrated that \cii\ is spatially extended compared with \oiii\ (e.g., \citealt{carniani2017, fujimoto2019, vallini2021, witstok2022}), we examined the spatial sizes of these emission lines in Mrk~54 with the surface brightness profiles. We found that \oiii\ is barely resolved with an upper limit of the intrinsic size of 8\farcs8, whereas \cii\ is only marginally resolved with an intrinsic size of 7\farcs9. Thus, we do not find a clear evidence of spatially extended \cii\ emission in Mrk~54 at the FIFI-LS's spatial resolution.

With {\it sospex}, we found that \cii\ and \oiii\ have the line fluxes of $(4.68 \pm 0.16)$ and $(1.01 \pm 0.27)\times10^{-16}$~W~m$^{-2}$, FWHMs of $329 \pm 13$ and $292 \pm 90$~km~s$^{-1}$, and redshifts of $0.04497 \pm 0.00002$ and $0.0443 \pm 0.0001$, respectively. The \cii\ and \oiii\ redshift is consistent with that of the optical redshift, $0.0447 \pm 0.0002$ (\citealt{deharveng2001}), within $1\sigma$ and $2\sigma$ uncertainties, respectively. The corresponding \cii\ and \oiii\ luminosities are $(5.90 \pm 0.21)$ and $(1.27 \pm 0.34)\times 10^{8}~L_{\rm \odot}$, respectively. 

\subsection{Comparisons of Line Luminosities to the Literature Samples} \label{subsec:sofia_obs_comparisons}

The left (right) panel of Figure \ref{fig:Lline_Lratio} compares the FIR and \cii\ (\oiii) luminosity of Mrk~54 with a compiled sample of the literature, where the FIR luminosity of Mrk~54 is obtained from photometry with {\it IRAS} 60 and 100~\micron\ as in  \cite{helou1985}. The literature sample first includes the data points of \cite{herrera-camus2018}, a compilation of local {\sc H ii} galaxies, LINER, Seyfert, and high-$z$ galaxies. The literature data also include a sub-sample of the Lyman Alpha Reference Sample (LARS; \citealt{hayes2013}), $z\sim0.03-0.18$ galaxies characterized by its ongoing starburst and Ly$\alpha$ emission. Seven and two of the LARS galaxies were observed in \cii\ and \oiii, respectively (\citealt{puschnig2020}). Finally, we also include the literature data points of \cite{Smirnova-Pinchukova2019}, a sub-sample of five nearby luminous Seyfert 1 AGN host galaxies observed in \cii. Interestingly, Mrk~54 has strong \cii\ emission that accounts for as high as $\sim1$\% of the FIR luminosity, whereas it has only moderate \oiii\ emission.  Based on the two line luminosities, we obtain \oiii/\cii~$=0.22\pm0.06$, much lower than those obtained in another LCE, Haro 11 ($\sim2$) and high-$z$ ALMA galaxies ($\sim3-10$).

Strong \cii\ emission in Mrk~54 could be due to the presence of outflow as discussed in e.g., \cite{Smirnova-Pinchukova2019} and \cite{puschnig2020}. Alternatively, previous theoretical works have shown that FIR line luminosity ratios can be affected by the presence of X-ray Dominated Regions in AGNs (e.g., \citealt{meijerink2007, wolfire2022}). Nevertheless, observational studies of \cite{decarli2022} and \cite{walter2018} have demonstrated that the FIR line ratios can be reasonably reproduced by the photo-dissociated region models, indicating that the presence of hidden AGN activity in Mrk~54, if any, may not be the primary cause of the strong \cii\ emission.

\subsection{Spatial and Velocity Offsets of \cii\ and \oiii\ in Mrk~54} \label{subsec:sofia_obs_offset}

Figure \ref{fig:oiii_cii_panstarrs} shows a direct comparison of the spatial distributions of \oiii, \cii\, and the $i$-band position of Mrk~54 in the FoV of \oiii, where the contours are drawn to show the flux centroids. We find that the \cii\ flux centroid matches the $i$-band position of Mrk~54, whereas the \oiii\ flux centroid is spatially offset by $\sim 5\arcsec$ ($\sim 4$~kpc). In the case of weak detections, the centroid of a point source is not well defined since almost no flux is detected in the adjacent pixels. Nevertheless, the position of the center of the emission can be reconstructed using spatial dithering as done in our observations. Another possibile cause of mismatch is a misalignment of the blue and red arrays, since the two spectra are acquired simultaneously. This is possible only if the rotation of the K-mirror of FIFI-LS~\citep[see][]{colditz2020}, the beam rotator that rotates the instrument's field of view counteracting the sky rotation experienced by the SOFIA telescope, is not accurately taken into account during the observation. The K-mirror parameters are measured before each FIFI-LS series since the cooling of the instrument slightly modifies its optical path. Since there are no similar issues with other observations in the two series when Mrk~54 was observed we can also exclude this possibility.  Finally, we notice that the peak of the \oiii\ emission is detected in the same position in the two independent observations obtained in the two different FIFI-LS series. This strongly suggests that the detection is real as well as the difference in position with respect to the peak of the \cii\ emission.

Interestingly, similar spatial offsets among the \cii, \oiii, and rest-frame ultraviolet continuum emitting regions were observed in galaxies in the EoR; \cite{carniani2017} showed a spatial offset in each of \cii, \oiii, and the rest-frame ultraviolet stellar continuum in a $z=7.11$ galaxy, BDF-3229 (see also \citealt{laporte2019}). More specifically, the galaxy shows a 4~kpc offset between the ultraviolet stellar continuum and \cii\ regions, and $1 - 30$~kpc offset between the \cii\ and \oiii\ emitting regions\footnote{The BDF-3299 galaxy has three clumps of \oiii, whose spatial offset from the \cii\ emission varies from $1$ to $30$~kpc.}. Furthermore, in the local Universe, \cite{cigan2016} have shown a spatial offset of \oiii\ and \cii\ in local metal poor dwarf galaxies, DDO 69 and DDO 70, although the offsets are as small as 0.1 - 0.2 kpc. 
On the theoretical viewpoint, based on cosmological hydrodynamic simulations combined with radiative transfer calculations, \cite{katz2019} have shown that the spatial offsets of \cii\ and \oiii\ can be caused by a clumpy structure due to e.g., merger events, presence of small satellites or a fragmented disc (see also \citealt{vallini2017} for another explanation of the spatial offset due to the stellar photo-evaporation feedback effect). Similarly, based on cosmological hydrodynamic simulations implementing the photoionization models of CLOUDY, \cite{moriwaki2018} have shown that there can be a spatial offset between \oiii\ 88 \micron\ and the rest-frame optical continuum that traces the distribution of the bulk stellar population (see their Fig. 3). The spatial offset could be due to the fact that the rest-frame optical continuum traces the older stellar population, whereas the \oiii\ 88 \micron\ traces the outer region of the actively star-forming regions (private communication with K. Moriwaki). These studies support the claim that the \oiii\ emission in Mrk~54 could be real. Hereafter, we treat the \oiii\ detection as real. We stress that the \oiii/\cii\ ratio in Mrk~54 remains low even if \oiii\ is considered undetected. 

Finally, we note that the \cii\  line is redshifted with respect to the \oiii\ line by $192\pm30$ km s$^{-1}$ in Mrk~54 (Fig. \ref{fig:mrk54}). At high-$z$,  \cite{carniani2017} have reported that \cii\ is blue-shifted with respect to \oiii\ by $\sim500$ km s$^{-1}$ in the BDF-3299 galaxy at $z=7.2$ (see also, e.g., \citealt{algera2023} for other high-$z$ examples). The authors have ascribed the velocity offset to a different kinematics of \cii- and \oiii-emitting gas, i.e., predominantly-neutral and the ionized gas phase, respectively. Because Mrk~54 is a LCE, such a different kinematics of gas might indicate a hint of LyC photon leakage from galaxies into the surrounding IGM. However, we leave further discussion to future papers because another local LCE, Haro~11, does not show any velocity offset between the \cii\ and \oiii\ lines (\citealt{cormier2015}).

\section{Literature Samples} \label{sec:literature_sample}

For our analysis we make use of data from HDGS and LT. The HDGS sample consists of 50 local metal-poor dwarf galaxies including the LCE Haro~11 (\citealt{cormier2012}) and it is characterized by low gas-phase metallicity ranging from near solar metallicity down to $12+\log($O/H$)$ = 7.14. In most cases, these galaxies have PACS spectroscopic observations for both \cii~158~\micron\ and \oiii~88~\micron, followed by \oi~63~\micron\ and weaker lines of \oi~145~\micron\ and \nii~122/205~\micron, in addition to PACS imaging. The sample also includes measurements of various ancillary multi-wavelength photometric (UV to radio) and spectroscopic data, best suitable for examining the correlation between the \oiii/\cii\ line luminosity ratios and other observables or physical quantities. Specifically, we examine the correlations against (i) the ionization parameter, $U_{\rm ion}$, traced by $\rm O_{32}$ $\equiv$ \oiii~5007~\AA/\oii~3727~\AA, (ii) specific SFR defined as the SFR per unit stellar mass, sSFR, (iii) dust temperature, $T_{\rm dust}$,  (iv) ionized-to-neutral gas fraction traced by the \oiii\ 88 \micron-to-\oi\ 63 \micron\ line ratio, and (v) the gas-phase metallicity in units of 12+log(O/H). The choice of these parameters is motivated by theoretical predictions of \cite{vallini2021} and \cite{katz2022} except for $T_{\rm d}$. The correlation against $T_{\rm d}$ is considered because \cite{walter2018} have shown a positive correlation between \oiii/\cii\ and $T_{\rm d}$ (see Fig.~4 in their paper).  

From the parent sample of 50 galaxies, we first narrow down the sample to 41 galaxies with both \cii\ 158 \micron\ and \oiii\ 88 \micron\ data across the entire galaxy scale. Among these, the numbers of galaxies with $\rm O_{32}$, sSFR, $T_{\rm dust}$, FIR \oiii/\oi, and $12+\log($O/H$)$ are 18, 36, 33, 35, and 41, respectively. The stellar masses and SFRs are adopted from \cite{madden2013} and \cite{de_looze2014}, respectively\footnote{Note that \cite{madden2013} also reported SFRs estimated from the IR luminosity. 
Because the sample is expected to be dust-poor, this method can underestimate the SFRs. We thus use the estimates in \cite{de_looze2014} from {\it GALEX} far ultraviolet and {\it Spitzer}/MIPS 24 \micron, i.e., SFR(UV+IR). Because \cite{de_looze2014} obtained SFR(UV+IR) for a sub-sample of the HDGS, we first obtained the relation between SFR(UV) and SFR(UV+IR) in the sub-sample, and applied a typical correction to the rest of the sample.}. 
The gas-phase metallicities of the sample are adopted from \cite{madden2013} and references therein, which are estimated from optical lines. We compute $\rm O_{32}$ in the sample using these literature data\footnote{The optical line spectroscopy of the sample galaxies were often performed with slit modes around bright star clusters. For fairer comparisons of multi-wavelength spectroscopic data, optical integral field unit spectroscopy will be crucial. Note that previous studies that use the HDGS's FIR and optical data also suffer from the same issue (\citealt{vallini2021, witstok2022}).}. Finally, we adopt $T_{\rm d}$ of the sample in \cite{Remy2013} estimated from multiple dust photometry using modified blackbody models\footnote{\cite{sommovigo2021} have introduced another technique to estimate $T_{\rm d}$ from the measurements of the \cii\ luminosity and continuum flux density at 158 \micron. Although we adopt $T_{\rm d}$ from SED fitting, we expect a similar correlation as the two methods provide consistent dust temperatures (Fig. 2 of \citealt{sommovigo2021}).}. The sub-sample used in this study has median values of $\rm O_{32}$ $=5.5$, sSFR $= 0.5$ Gyr$^{-1}$ , $T_{\rm d} = 32$ K, FIR \oiii/\oi\ $=2.7$, and $12+\log($O/H$)$ = 8.0. 

The galaxies in the LT survey share similar properties to the HDGS galaxies such as low gas-phase metallicity, but they are characterized by lower surface brightness and moderate star formation activity. Four LT galaxies in \cite{cigan2016} were observed in both \cii\ and \oiii, and have lower \oiii/\cii\ luminosity ratios than the HDGS galaxies (\citealt{cigan2016}). Thus, the sample is useful to accurately examine the correlation between \oiii/\cii\ and 12+log(O/H). The numbers of galaxies with $\rm O_{32}$, sSFR, $T_{\rm dust}$, FIR \oiii/\oi\, and $12+\log($O/H$)$ are 0, 2, 1, 2, and 4, respectively, and the values are adopted from \cite{cigan2016} and \cite{cigan2021}.

\section{Characteristics of High \oiii-to-\cii\ Galaxies} \label{sec:comparison_samples}

\begin{figure*}[]
\begin{center}
\hspace*{-0.5cm}
\includegraphics[width=18cm]{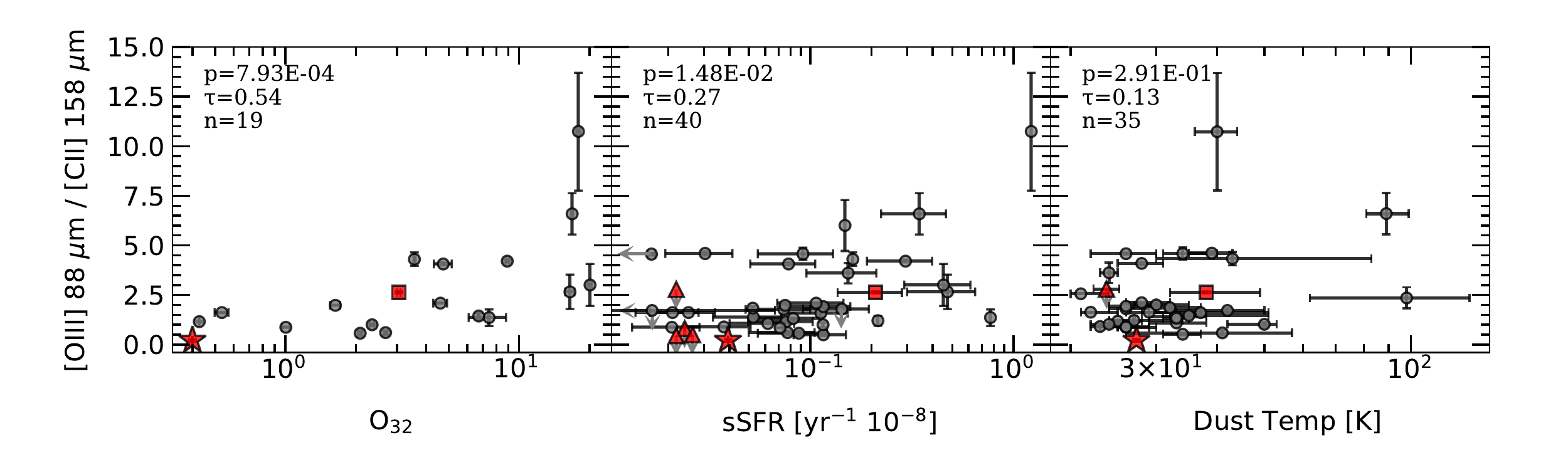}
\includegraphics[width=14cm]{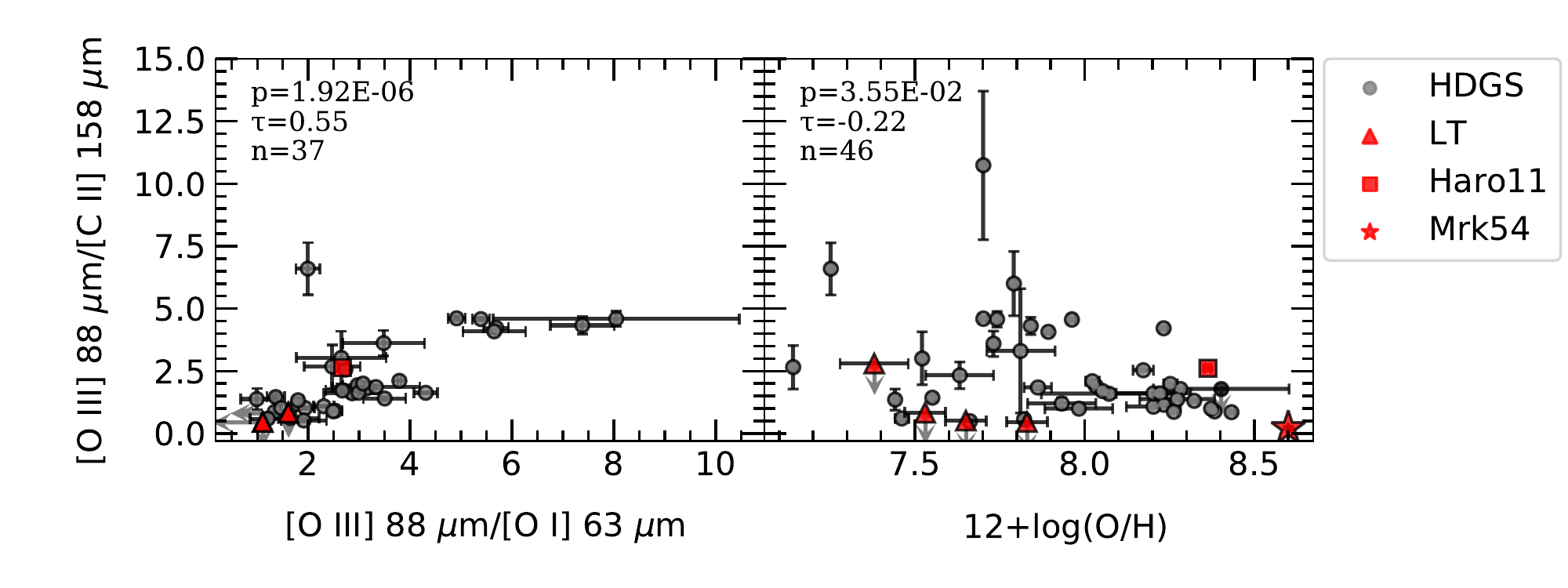}
\vspace*{-0.5cm}
\end{center}
\caption{
Correlations of \oiii/\cii\ line luminosity ratios plotted against $\rm O_{32}$, sSFR, $T_{\rm d}$, the FIR \oiii/\oi\ luminosity ratios, and gas phase metallicity in units of 12+log(O/H). In each panel, the red star and square indicate Mrk~54 and Haro~11, respectively. Gray circles show the HDGS galaxies, whereas the red triangles show the LT galaxies. Note that Mrk~54 does not have a measurement of the FIR \oiii/\oi\ line ratio, whereas the LT galaxies do not have measurements of $\rm O_{32}$. In the upper middle panel, the sSFR values of the LT galaxies are shifted up to 0.1 dex for the display purpose. In each panel, the values in the top left corners are the $p$-value, Kendall's correlation coefficient ($\tau$), and number of galaxies (n) used to examine the correlations.
\label{fig:OIII_CII_correlations}}
\end{figure*}

We examined the correlation of the \oiii/\cii\ line luminosity ratios with other observable quantities using our combined sample which consists of Mrk~54, HDGS galaxies including Haro~11, and LT galaxies. We use the Kendall rank correlation test to statistically examine the correlations, where a $p$-value of  0.05 is used to reject the null hypothesis that the two quantities do not correlate. The correlations and relative Kendall's correlation coefficients are plotted in Figure~\ref{fig:OIII_CII_correlations}.

The results show that objects with high \oiii/\cii\ line luminosity ratios are characterized by high $\rm O_{32}$ (i.e., high $U_{\rm ion}$), high sSFR (i.e., undergoing a bursty event or a young stellar age), high FIR \oiii/\oi\ (i.e., large volume fraction of ionized gas to neutral gas), and low metallicity. The result that high \oiii/\cii\ shows high ionization parameters and burstiness is consistent with \cite{harikane2020}, \cite{Sugahara2022}, and \cite{algera2023}, respectively. We do not observe a strong correlation between \oiii/\cii\ and $T_{\rm d}$ (but see \citealt{walter2018, algera2023}). The strongest correlations are found in $\rm O_{32}$ (i.e., $U_{\rm ion}$) and high FIR \oiii/\oi\ (i.e., the ionized-to-neutral gas volume filling factor), which are in qualitative agreements with the claims of \cite{harikane2020} and \cite{katz2022}. Although we found a correlation between \oiii/\cii\ and metallicity, it should be noted that all the four LT galaxies have low \oiii/\cii\ ratios despite their low metallicity, supporting the claims of \cite{vallini2021} who showed that the metallicity is not a main factor that changes \oiii/\cii. The correlations with metallicity and $T_{\rm d}$ have also been investigated in \cite{cigan2016}, but to our knowledge, this is the first observational study that examines the correlations with $\rm O_{32}$.

\section{Discussion} \label{sec:discussion}

\begin{figure*}[]
\begin{center}
\hspace*{-1cm}
\vspace*{-0.5cm}
\includegraphics[width=18cm]{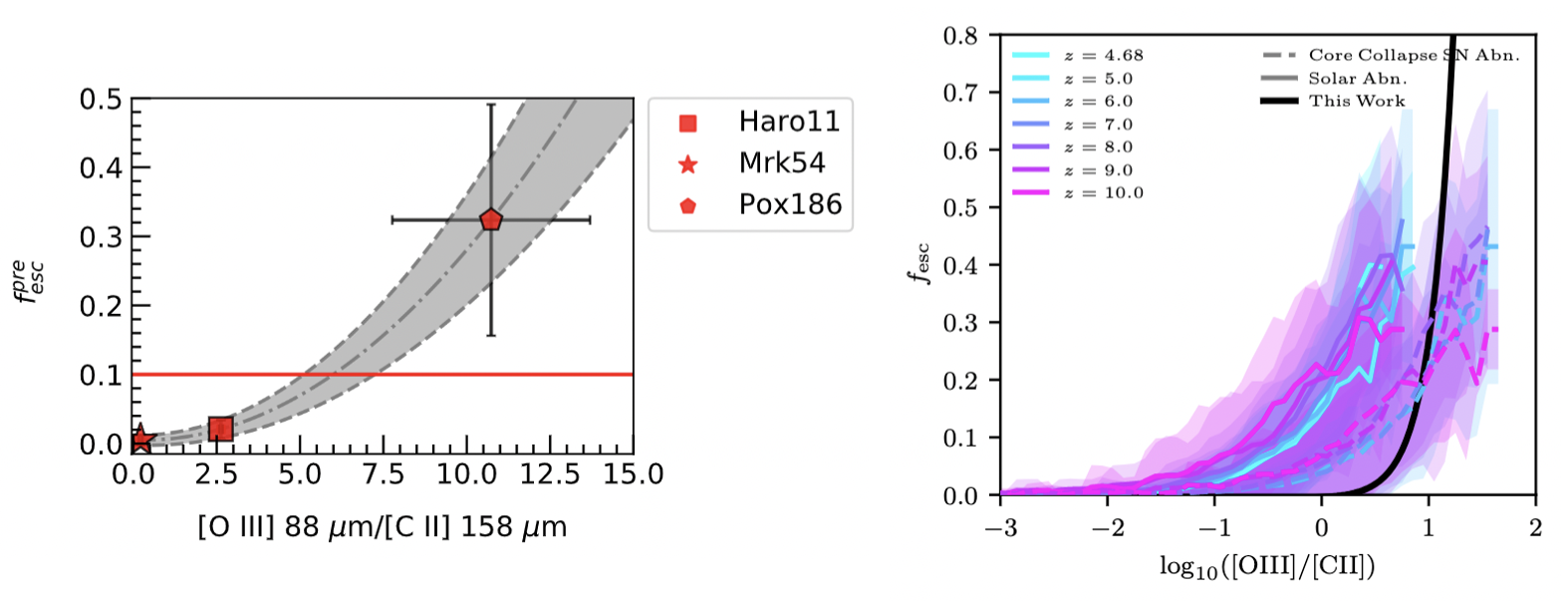}
\end{center}
\caption{
({\it Left}) Predicted value of the LyC escape fraction, $f_{\rm esc}^{pre}$, obtained from the \oiii/\cii\ luminosity ratio using Figure~\ref{fig:OIII_CII_correlations}. The shaded region indicates the $1\sigma$ uncertainty of the  \oiii/\cii-$f_{\rm esc}^{pre}$ relation. The red star and square indicate Mrk~54 and Haro~11, respectively, that have direct measurements of $f_{\rm esc}$. The red pentagon shows POX~186 with the highest \oiii/\cii\ luminosity ratio, $\sim10$, in the local Universe. 
({\it Right}) Comparisons of the best-fit   \oiii/\cii-$f_{\rm esc}^{pre}$ relation (black solid line) and the results of cosmological hydrodynamical simulations in \cite{katz2022} (colored shaded regions and lines). As in \cite{katz2022}, the colored lines and shaded region indicate the mean relation and $1\sigma$ scatter for different redshift bins. The solid and dashed lines represent the results with assumed values of Solar abundances and core-collapse supernovae abundances, respectively.
\label{fig:fesc}}
\end{figure*}

\subsection{\oiii/\cii-$f_{\rm esc}$ Relation } \label{subsec:discussion1}
As mentioned in \S \ref{sec:intro}, a quantitative description of the cosmic reionization process requires estimates of $f_{\rm esc}$ in galaxies in the EoR. We discuss the possibility to use the FIR \oiii/\cii\ luminosity ratios to infer $f_{\rm esc}$. 

Based on observations of LCEs and non-LCEs among the local galaxies, \cite{Chisholm2018} have derived an equation that links $\rm O_{32}$ to $f_{\rm esc}$ as follows, 
\begin{equation}
f_{\rm esc, O_{32}}^{\rm pre} = (0.0017 \pm 0.0004) {\rm O_{32}}^{2} + (0.005 \pm 0.007), 
\end{equation}
where $f_{\rm esc, O_{32}}^{\rm pre}$ is a predicted value of the LyC escape fraction from $\rm O_{32}$ (see also \citealt{izotov2016b}). Note that the correlation between $f_{\rm esc}$ and O$_{\rm 32}$ has been statistically confirmed in a recent study of \cite{flury2022b}, who significantly increases the sample of LCEs and non-LCEs. Because an equation relating the two quantities is not provided in \cite{flury2022b}, in this study, we use the relation of \cite{Chisholm2018} for simplicity. 
In Fig. \ref{fig:OIII_CII_correlations}, we have obtained a positive correlation between optical $\rm O_{32}$ and FIR \oiii/\cii. If we fit the correlation with a linear function, it can be written as 
\begin{equation}
{\rm O_{32} = (1.311 \pm 0.010) [O III]/[C II] + (-0.384 \pm 0.016)},  
\end{equation}
using a package of {\tt scipy.optimize.curve{\_}fit}.
Combining these two equations, we obtain the following equation that relates \oiii/\cii\ to $f_{\rm esc}$, 
\begin{eqnarray}
f_{\rm esc}^{\rm pre} = (0.0029 \pm 0.0007) ({\rm [OIII]/[CII]})^{2} + \nonumber \\
(-0.0017 \pm 0.0004) {\rm [OIII]/[CII]} + (-0.005 \pm 0.007). \nonumber \\
\end{eqnarray}
The positive correlation between the two quantities is in qualitative agreement with the predictions of \cite{inoue2016}, \cite{katz2022}, and \cite{ramabason2022}.

In the left panel of Figure \ref{fig:fesc}, we plot the predicted relation of LyC escape fraction along with the data points of Mrk~54, Haro~11, and POX~186. Mrk~54 and Haro~11 have direct measurements of the LyC escape fraction. Mrk~54 and Haro~11 have the measured escape fractions of  $0.025\pm0.007$ (\citealt{leitherer2016}) and $0.033\pm0.007$ (\citealt{leitet2011}), whereas the predicted values are $f_{\rm esc}^{\rm pre} = $ $0.005^{+0.007}_{-0.005}$ and $0.021 \pm 0.013$, respectively. The measured and predicted values in Mrk~54 and Haro 11 are consistent with each other within $2\sigma$ and $1\sigma$ uncertainties. 

In the left panel of Fig. \ref{fig:fesc}, a galaxy called POX~186 shows the highest \oiii/\cii\ luminosity ratio $\sim10$ among the local galaxies, which corresponds to a high $f^{\rm pre}_{\rm esc} \sim(0.32\pm0.15)$. Although direct LyC observations are not performed in the galaxy, its intense star formation activity (sSFR $\sim 12$ Gyr$^{-1}$) also supports that the galaxy is a candidate strong LCE. Recently, \cite{eggen2021} have performed integral field unit observations of optical emission lines, and have found the evidence of strong ionized-gas outflow that can facilitate the LyC escape. The indirect signature of strong LyC escape in POX~186 supports that the FIR \oiii/\cii\ ratios can be used at least to search for promising LCE candidates. 

\subsection{Implications for High Redshift ALMA Studies} \label{subsec:discussion2}

We now turn our attention to the results of high \oiii/\cii\ luminosity ratios in the UV-selected galaxies at $z\sim6-9$ (e.g., \citealt{hashimoto2019a, bakx2020, harikane2020, witstok2022}). In the following, we adopt the latest line luminosity measurements of nine ALMA \oiii\ 88 \micron\ galaxies at $z\sim6-9$ from \cite{carniani2020}. The nine galaxies were detected in both emission lines, and have \oiii/\cii\ $\sim 3-10$\footnote{Among the nine high-$z$ galaxies, \cii\ has been originally reported to be undetected in SXDF-NB1006-2 (\citealt{inoue2016}), A2744-YD4 (\citealt{laporte2019}), and MACS1149-JD1 (\cite{laporte2019}). Later, \cite{carniani2020} have reanalyzed the \cii\ data and found $4\sigma$ detections.}. The only exception is SXDF-NB1006-2 where we adopt a new measurement of \oiii/\cii\ $= 6.1 - 9.6$ (\citealt{ren2023}).

Under the assumption that the same \oiii/\cii-$f_{\rm esc}$ relation can be applied to these EoR galaxies, the nine high-$z$ ALMA \oiii\ 88 \micron\ galaxies are expected to have $f_{\rm esc}\sim 0.01-0.25$. In particular, MACS 0416\_Y1 at $z=8.31$ (\citealt{tamura2019, bakx2020}), J0235 at $z=6.0$ (\citealt{harikane2020}) and SXDF-NB1006-2 at $z=7.21$ (\citealt{inoue2016, ren2023}) have \oiii/\cii\ $\gtrsim7$, and could be candidate LCEs with $f_{\rm esc}\gtrsim0.1$, which is an average value required for galaxies to explain the cosmic reionization process (e.g., \citealt{inoue2006, finkelstein2019, ma2020}).  

Based on cosmological hydrodynamical simulations, \cite{katz2022} have obtained a relation between $f_{\rm esc}^{\rm pre}$ and \oiii/\cii\ at $z\sim4-10$. The right panel of Figure \ref{fig:fesc} shows comparisons of our eq. (3) and the results of cosmological hydrodynamical simulations in \cite{katz2022} for mock galaxies with SFR $>10^{-2}$~$M_{\rm \odot}$~yr$^{-1}$. If we limit the luminosity ratio range to the observed one of log(\oiii/\cii)$\sim0-1$, we find that our best-fit relation systematically predict lower $f_{\rm esc}$ values. \cite{katz2022} have also presented the $f_{\rm esc}^{\rm pre}$-\oiii/\cii\ relation taking the mass dependence of $f_{\rm esc}^{\rm pre}$ into account (see their eq. 4). With this relation,  the authors have obtained  escape fractions of $0.005 - 0.022$ in the same nine galaxies at $z=6-9$, even lower than our predictions. These results imply that we also need to take the mass dependence of $f_{\rm esc}^{\rm pre}$ into account. We do note, however, that other observational techniques (e.g., the peak separation of Ly$\alpha$ and $\rm O_{32}$ etc) usually do not take the mass dependence into account either. As the number of LCEs with \oiii/\cii\ is just two, it is impossible to observationally examine the mass dependence of the \oiii/\cii-$f_{\rm esc}$ relation in a statistical manner. 
We also note that the stellar masses of high-$z$ galaxies are generally largely uncertain due to the lack of rest-frame optical data. Furthermore, there is observational evidence that massive galaxies do not necessarily have low $f_{\rm esc}$. Indeed, \cite{Marques-Chaves2021} and \cite{Marques-Chaves2022} have shown that extremely UV luminous and massive (log($M_{\rm *}/M_{\rm \odot}$) $\sim 9.9\pm0.1$) galaxies have significant LyC escape fractions of $\sim 20-90$\%. 
Taking these into account, we do not regard the different $f_{\rm esc}$ values as a serious tension.

\subsection{Limitation of This Study and Future Prospects} \label{subsec:discussion3}

We discuss caveats and limitations of the relation between \oiii/\cii\ and $f_{\rm esc}$, then discuss how we can overcome the issues in the future. First, we derived an empirical relation between \oiii/\cii\ and $f_{\rm esc}$ using (1) the known relation of $f_{\rm esc}$ and $\rm O_{32}$ in \cite{Chisholm2018}, and (2) the correlation of \oiii/\cii\ and $\rm O_{32}$ based on the combined sample of HDGS, LTS, and Mrk~54. Although the relation of $f_{\rm esc}$ and $\rm O_{32}$ has been statistically confirmed in the latest study of \cite{flury2022b}, there exists a large scatter in the relation, not accurately traced by eq. (1). Secondly, the relation between \oiii/\cii\ and $f_{\rm esc}$ can be severely affected by different selection functions adopted in the samples of local LCEs/non-LCEs in \cite{Chisholm2018} and the HDGS/LTS galaxies. Because there are only two known LCEs with the FIR line observations, Haro~11 and Mrk~54, however, we stress that the present study offers the best effort to empirically calibrate the relation of \oiii/\cii\ and $f_{\rm esc}$.

Clearly, a larger number of LCEs with FIR line observations is needed to refine the \oiii/\cii-$f_{\rm esc}$ relation. Unfortunately, because of the closure of SOFIA, this kind of observations will be only possible with future FIR probes such as the PRobe Infrared Mission for Astrophysics (PRIMA) which will be able to target both \cii\ and \oiii\ up to $z=0.27$. Because some nearby LCEs at $z\sim0.2-0.4$  have  $f_{\rm esc}$ well beyond 10\%, in contrast to the low values of Haro~11 and Mrk~54 ($f_{\rm esc}\sim2-3$\%), such observations will be crucial to calibrate the  \oiii/\cii-$f_{\rm esc}$ relation up to the high $f_{\rm esc}$ regime. A better calibrated relationship could be therefore a new powerful diagnostic tool to search for LCE candidates in addition to the current rest-frame UV/optical techniques. 

Finally, we have assumed that the same relation between $f_{\rm esc}$ and \oiii/\cii\ can be applied to high-$z$ galaxies in \S \ref{subsec:discussion2}. Although we do not have direct observations of rest-frame optical emission lines of these ALMA high-$z$ galaxies, the {\it James Webb Space Telescope} will provide rest-frame optical emission lines of ALMA \oiii\ 88 \micron\ emitters at $z\sim 6-9$ (e.g., GO1 \#1840, PIs: J. {\'A}lvarez-M{\'a}rquez and T. Hashimoto as well as GTO programs of \#1776 PI: R. Windhorst, \#1199 PI: M. Stiavelli, \#1262 PI: N. Luetzgendorf, and \#1264 PI: L. Colina). With these data, it will be possible to test if the \oiii/\cii-O$_{\rm 32}$ relation holds at $z\sim 6-9$.

\section{Summary} \label{sec:summary}

We presented SOFIA \cii~158~\micron\ and \oiii~88~\micron\ observations of the local LCE, Mrk~54, at the luminosity distance of $\sim$191 Mpc. This is only the second LCE, after Haro~11, observed in the FIR lines, offering the opportunity to test if the \oiii/\cii\ line luminosity ratio can be used as a tracer of $f_{\rm esc}$, as proposed by \cite{inoue2016} and \cite{katz2020}. Interestingly, we find that Mrk~54 has strong \cii\ emission, whereas it has only moderate \oiii\ emission, resulting in the low \oiii/\cii\ luminosity ratio of $0.22\pm0.06$. Combining Mrk~54 with the literature samples of the HDGS and LT surveys, we find that the \oiii/\cii\ luminosity ratio correlates with $\rm O_{32}$, sSFR, FIR \oiii/\oi, and anti-correlates with the gas-phase metallicity in units of 12+log(O/H). The strongest correlations are found with $\rm O_{32}$ and FIR \oiii/\oi. By combining the analytical form of $f_{\rm esc}$ and $\rm O_{32}$ in \cite{Chisholm2018} and the correlation between \oiii/\cii\ and $\rm O_{32}$ found in this study, we obtain a relation between $f_{\rm esc}$ and \oiii/\cii. The two confirmed LCEs, Haro~11 and Mrk~54, roughly follow the \oiii/\cii-$f_{\rm esc}$ relation within uncertainties of $1\sigma$ and $2\sigma$, respectively. If we assume that the same \oiii/\cii-$f_{\rm esc}$ relation holds at high-$z$, galaxies with \oiii/\cii\ $\gtrsim7$ could have $f_{\rm esc} \gtrsim10$\%, significantly contributing to the reionization process. The presented multi-wavelength study will serve as an invaluable reference for ongoing and planned {\it JWST} observations of ALMA \oiii\ 88 \micron\ emitters at high-$z$.


\acknowledgments
We thank an anonymous referee for valuable comments that have greatly improved the paper. We are grateful to Rodrigo Herrera-Camus, Harley Katz, and Ren W. Yi for providing us with their data. 
We appreciate Christian Fischer, Irina Smirnova-Pinchukova, Harley Katz, and Kana Moriwaki for useful discussion.
This research is based on data from the SOFIA Observatory, jointly operated by USRA (under NASA contract NNA17BF53C) and DSI (under DLR contract 50 OK 0901 to the Stuttgart University). TH was supported by Leading Initiative for Excellent Young Researchers, MEXT, Japan (HJH02007) and by JSPS KAKENHI Grant Numbers (20K22358 and 22H01258). 
AKI, YS, and YF are supported by NAOJ ALMA Scientific Research Grant Numbers 2020-16B.
EZ acknowledges funding from the Swedish National Space Agency. YT was supported by JSPS KAKENHI Grant Number (22H04939) and by NAOJ ALMA Scientific Research Grant Number 2018-18B. MH is fellow of the Knut \& Alice Wallenberg Foundation. YH was supported by JSPS KAKENHI Grant Number (21H04489) and JST FOREST Program, Grant Number (JP-MJFR202Z).

\vspace{5mm}
\facilities{SOFIA (FIFI-LS), IRAS}

\software{
Astropy \citep{astropy2013,astropy2018}, SciPy \citep{Virtanen2020} and sospex \citep[\url{www.github.com/darioflute/sospex,}][]{fadda2018}
}

\clearpage


\end{document}